\documentclass[twocolumn,twoside,slac_two]{revtex4}
\usepackage{graphicx,euscript}
\usepackage{fancyhdr}
\def\F{{\EuScript F}}
\def\e{\varepsilon}

\begin{document}

\title{\boldmath Inclusive semileptonic decays of the $B$ meson}

%

\author{Bjorn O. Lange}
\affiliation{Center for Theoretical Physics, MIT, Cambridge, MA, U.S.A.}

\begin{abstract}
This talk is a short review on the theoretical issues and
uncertainties in the calculation of partial decay rates in inclusive
$B$ decays. The main emphasis is on charmless semileptonic decays,
the $\bar B \to X_s\gamma$ photon spectrum, and the extraction of
$|V_{ub}|$ using model-independent methods. 
\\[3mm]
Preprint: MIT-CTP 3741
\end{abstract}

\maketitle

\thispagestyle{fancy}


\section{Introduction}

Over-constraining the unitarity triangle using many different physical
processes is a crucial test of the CKM picture of {$\cal CP$}
violation and the validity of the Standard Model. The topic of this
talk is the theoretical framework and uncertainty estimation for
partial rates in $\bar B\to X_u\,l^-\bar\nu$ decays and the extraction
of the element $V_{ub}$. Its magnitude is proportional to the length
of the side opposite the well-measured angle $\beta$ of the unitarity
triangle.

Because of an overwhelming background from $\bar B\to X_c\,l^-\bar\nu$
decays in a large portion of the phase space, experimental cuts are
employed which suppress such background events. The bulk of this talk
deals with the theoretical techniques used for calculating the
surviving partial decay rates.  We may identify two regions of
phase-space for partial rates according to the integration domains in
the hadronic variables
\begin{equation}
P_\pm = E_X \mp |\vec P_X|\;,
\end{equation}
where $E_X$ and $\vec P_X$ are the energy and 3-momentum of the
hadronic final state in the $B$ meson rest frame. The available phase
space in these variables is $M_\pi^2/P_- \le P_+ \le M_B-2 E_l \le P_-
\le M_B$, where $M_\pi^2$ is the invariant mass of the lightest
possible hadronic final state, i.e.\ the pion mass squared, and $E_l$
is the energy of the charged lepton $l^-$. For the rest of this talk
we will neglect the finite pion mass for simplicity. Note that the
product $P_+ P_- = M_X^2$ gives the invariant mass of the hadronic
state. Hence the charmed background is eliminated as long as the phase
space $P_+ P_- \ge M_D^2$ is cut out. This is achieved in various
ways, e.g.\ by imposing restrictions on the leptonic invariant mass
$q^2 = (M_B-P_+)(M_B-P_-)$, or $E_l$, or $P_+$, or $M_X$, or
combinations of them. The theoretical tools used for the calculation
of such partial rates depend on whether the range of integrations are
such that both $P_+$ and $P_-$ are much larger than $\Lambda_{\rm
QCD}$ (operator product expansion, OPE), or $P_+$ of order
$\Lambda_{\rm QCD}$ while $P_-$ much larger (QCD factorization
theorems in this ``shape-function region''). If both $P_+$ and $P_-$
are small then resonances are not sufficiently smeared and the
calculations are unreliable due to a breakdown of quark-hadron
duality.

It is convenient to use a set of dimensionless variables which we
define as $y=(P_--P_+)/(M_B-P_+)$ and $\e = 1 - 2 E_l/(M_B-P_+)$, for
which the phase-space is simply $0 \le \e \le y \le 1$. The dependence
of the fully differential decay rate on the charged lepton energy --
or on $\e$ in the dimensionless variables -- can be separated from the
dependence on hadronic variables, and the differential decay rate
reads \cite{Lange:2005yw}
\begin{eqnarray}
\frac{d^3\Gamma_u}{dP_+ dy\,d\e} &=& \frac{G_F^2 |V_{ub}|^2}{16\pi^3}
U_y(\mu_h,\mu_i) \, (M_B-P_+)^5 \nonumber \\
&& \Big[ (y-\e)(1-y+\e)\, \F_1 \nonumber \\
&& \mbox{} + y (1-y)\, \F_2 + \e (y-\e)\, \F_3 \Big]\;, \nonumber 
\end{eqnarray}
where the structure functions $\F_i$ depend on both the hadronic
variables $P_+$ and $y$ and on the factorization scales $\mu_h$ and
$\mu_i$, but not on $\e$. It is worth noting that the above formula is
written entirely in terms of hadronic quantities, so that at this step
no uncertainty associated with our inability to determine the
$b$-quark mass $m_b$ is introduced.

The differential decay rate is formally independent of the
factorization scales $\mu_h$ and $\mu_i$. The scale dependence of the
structure functions $\F_i$ cancels against the evolution factor
$U_y(\mu_h,\mu_i)$, which sums (double) logarithms of the ratio of
both scales and carries also dependence on $y$. The ``natural''
scaling for the scales, that avoids large logarithms, is such that
$\mu_i^2 \sim \langle P_+ P_- \rangle$ and $\mu_h \sim \langle P_-
\rangle$, where the brackets indicate an average over the considered
region in phase space when integrating the differential decay rate. In
most applications the phase space surviving the experimental cut is
such that the shape-function region, where QCD factorization theorems
hold, is included and dominates the event rate. However, it is
desirable to have a description of the decay rate in the entire phase
space for e.g.\ studies of detector resolution effects. In the next
part of this talk we will first address the theoretical issues of the
differential decay rate in the shape-function region, and discuss the
non-trivial transition into the OPE region. The second part deals with
relations between semileptonic decays and the radiative $\bar B \to
X_s \gamma$ photon spectrum, which are constructed in such a way to be
model-independent at leading power.

\section{Factorization}\label{sec:2}

In the shape-function region there are three relevant scales in the
problem: the hard scale $\mu_h \sim \langle P_- \rangle \sim m_b$, the
intermediate scale $\mu_i^2 \sim \langle P_+ P_- \rangle \sim m_b
\Lambda_{\rm QCD}$, and the non-perturbative scale $\Lambda_{\rm
QCD}$. When systematically separating the physics effects at these
scales, the structures $\F_i$ factorize into three parts
correspondingly, the hard function, the jet function, and the shape
function \cite{Korchemsky:1994jb}.  In modern language this
factorization is achieved via a series of effective-field theory
calculations QCD$\to$SCET$\to$HQET. The matching coefficients are by
construction free of infra-red physics and capture the physics at the
hard and jet scale \cite{Bosch:2004th,Bauer:2003pi}. Symbolically we
can write $\F_i = H_i(y,\mu_h) J(y, P_+,\mu_i) \otimes \hat
S(\mu_i)+\ldots$, where $\otimes$ denotes a convolution integral, and
the ellipsis denote terms that are suppressed by $\Lambda_{\rm
QCD}/m_b$ as compared with the leading term. It was shown explicitly
in \cite{Bosch:2004cb, Lee:2004ja} that these power corrections
factorize themselves in a similar fashion. The major difference is,
however, that the leading-power shape function is unique, while at
subleading order more than one soft function appear. 

Let us return to the leading-power formula. Not only is the shape
function universal, but in fact also the jet function, and we may
combine the hard functions with the kinematic prefactors to write
\begin{eqnarray} \label{eq:LPfact}
\frac{d^3\Gamma_u^{(0)}}{dP_+ dy\,d\e} &=& \frac{G_F^2 |V_{ub}|^2}{192\pi^3}
U_y \, (M_B-P_+)^5 H_u(y,\e) \\
&& \mbox{} \times \int_0^{P_+} \!\! d\hat\omega\, ym_bJ(ym_b(P_+-\hat\omega)) 
\hat S(\hat\omega)\;. \nonumber 
\end{eqnarray}
In the above expressions we have neglected the scale dependence, which
is as discussed above. The superscript $^{(0)}$ on the left-hand side
indicates that this formula is the leading power contribution
only. The hard and jet functions have been calculated at one-loop
order in \cite{Bosch:2004th,Bauer:2003pi}, and recently the two-loop
result for the jet function has become available
\cite{Becher:2006qw}. The shape function $\hat S(\hat\omega)$, on the
other hand, is not calculable in perturbation theory and must be
determined by other means. This is where the universality of the shape
function comes into play. Considering the well-measured $\bar B\to
X_s\gamma$ photon spectrum near the endpoint of maximal photon energy
$E_\gamma$, one can state a formula similar to
equation~(\ref{eq:LPfact}), namely
\begin{eqnarray} \label{eq:LPfact2}
\frac{1}{\Gamma_s}\frac{d\Gamma_s^{(0)}}{dP_+} &=& \frac{U}{H_\Gamma}
\frac{(M_B-P_+)^3}{m_b^3} \\
&& \mbox{} \times \int_0^{P_+} \!\! d\hat\omega\, m_bJ(m_b(P_+-\hat\omega)) 
\hat S(\hat\omega)\;. \nonumber 
\end{eqnarray}
For later convenience we have normalized the photon spectrum to the
total decay rate. 

Many of the ingredients of this equation have already been
discussed. The main difference to the semileptonic decay is that a)
the short-distance physics is different leading to a different hard
function, and b) the phase space is given by $P_- = M_B$ (hence $y=1$)
and $P_+ = M_B-2 E_\gamma$. Taking this into account, $U$ still resums
logarithms of the ratio $\mu_h/\mu_i$, but no longer carries a
dependence\footnote{The relation is $U_y(\mu_h,\mu_i) = U(\mu_h,\mu_i)
\cdot y^{-2a_\Gamma(\mu_h,\mu_i)}$, where $a_\Gamma(\mu_h,\mu_i)$ is
the integrated cusp anomalous dimension. See \cite{Lange:2005yw} for
details.} on $y$. We encounter the same shape function and jet
function, with the latter evaluated with $y=1$.

The general idea for the determination of $|V_{ub}|$ from inclusive
$B$ decays is to use the $\bar B\to X_s\gamma$ photon spectrum to
extract the shape function, and subsequently use it for predictions of
arbitrary decay rates by integrating the differential decay rate
(\ref{eq:LPfact}). This is typically achieved by assuming a reasonable
functional form for the shape function in the domain where $\hat\omega
\sim \Lambda_{\rm QCD}$ with adjustable parameters, and fitting the
parameters to the data. The current state-of-the-art parameterizations
involve exponential-type, gaussian-type, or hyperbolic-type functions
with two free parameters, which can be linked to the calculation of
the first few moments of the shape function. Therefore the extraction
of the leading shape function is mainly an experimental issue, while
theory contributes via the calculation of moment constraints (where a
link to other decay processes is established, e.g.\ to $\bar B\to
X_c\,l^-\bar\nu$, see for example
\cite{Beneke:1999fe,Aubert:2004aw,Bauer:2004ve} and scheme
translations in \cite{Neubert:2004sp}), and via the explicit
calculation of the radiative tail of the shape function.

Beyond the leading-power approximation there are several corrections
that one should take into account. Non-perturbative (``hadronic'')
corrections are encoded in subleading shape functions
\cite{Bosch:2004cb, Lee:2004ja}, of which there are three new, independent
ones at tree level called $\hat t(\hat\omega), \hat u(\hat\omega)$ and
$\hat v(\hat\omega)$. Their functional form is unknown and the only
information available at present are their first few moments at
tree level. Furthermore there are corrections proportional to
arbitrary powers of the ratio of $P_+/m_b$, called ``kinematical''
corrections, which start at order ${\cal O}(\alpha_s)$ and come with
the leading shape function.

To make the extraction of the leading shape function from the $\bar
B\to X_s\gamma$ photon spectrum feasible, a certain combination of
$\hat t(\hat\omega), \hat u(\hat\omega)$ and $\hat v(\hat\omega)$ is
absorbed into the redefined leading shape function. The only structure
surviving at subleading power is then proportional to $(\bar\Lambda -
\hat\omega) \hat S(\hat\omega)$, where $\bar\Lambda = M_B-m_b$ is a
heavy-quark parameter.
It has recently been proposed to cross-check the findings by comparing
with $\bar B\to X_c\,l^-\bar\nu$ decay spectra, where a shape-function
region also exists \cite{Boos:2005qx}. However, so far many important
corrections -- for example subleading shape-function contributions --
have not been included.

\section{Transition into OPE region}

The region where an operator product expansion applies is reached when
the size of the integration domain over shape functions, i.e.\ the
maximal $P_+$, is much larger than $\Lambda_{\rm QCD}$.  The key
observation is that the factorization theorems for the shape-function
region connect with the traditional OPE calculation via {\em moments}
of the shape functions. More generally, if the integration domain
$\Delta$ in an integral of the form (with some function
$f(\hat\omega)$)
\begin{equation}
\int_0^{\Delta} d\hat\omega\; f(\hat\omega)\,\hat S(\hat\omega)
\end{equation}
is much larger than $\Lambda_{\rm QCD}$, then one can perform an
operator product expansion in the quantity $\Lambda_{\rm
QCD}/\bar\Delta$ \cite{Neubert:2004dd} with $\bar\Delta = \Delta -
\bar\Lambda$. An interpolation between the shape-function region and
the OPE region, and therefore a description for the entire phase
space, can now be achieved by essentially brute force. Let us
demonstrate the procedure for the partial decay rate with a cut on
$P_+$, i.e.
\begin{equation}
\Gamma_u(\Delta) = \int_0^{\Delta} dP_+ \frac{d\Gamma_u}{dP_+}\;.
\end{equation}
We will first consider the case of small $\Delta$, where the
factorization theorems are valid, and identify a certain contribution
(with label ``SF''). Then we will ask what happens to this
contribution when increasing $\Delta$ so that an OPE calculation is
valid (labeled ``OPE'').

The leading-power ``SF'' contribution (\ref{eq:LPfact}) feeds into all
powers in the OPE via the moments of the shape function, which involve
the heavy-quark parameters $m_b, \mu_\pi^2, \ldots$, but also misses
some contributions at every level in the power counting. We will
return to this last point shortly. The first subleading hadronic
``SF'' corrections (from subleading shape functions), which are known
at tree level, do not contribute to the leading-power ``OPE'' piece,
but again feed into all subleading powers below. Similarly the second
subleading hadronic ``SF'' corrections give no contribution to the
leading or the first subleading ``OPE'' piece. (Of course, when
expanding in $1/m_b$ there is no first-order power correction in the
OPE calculation.) Currently there is no complete categorization of
subsubleading shape functions at order $1/m_b^2$. However, it is via
the above observation that at least those which contribute to the
$1/m_b^2$ piece in the OPE calculation can be simulated
\cite{Lange:2005yw}.

As mentioned earlier, there are also kinematical radiative corrections
that are power suppressed in the shape-function region simply because
they are proportional to $\alpha_s(\bar\mu)\cdot(P_+/m_b)^k$, $k\ge
1$. Such ``SF'' contributions come with the leading shape function and
are promoted to leading power ``OPE'' terms when $P_+$ becomes of
order $m_b$. The exact ${\cal O}(\alpha_s)$ kinematic corrections are
known by comparing the factorized expressions with the fixed-order
partonic calculation in full QCD \cite{DeFazio:1999sv}.

By including all of the above contribution in the shape-function
region we correctly reproduce the OPE result when allowing $\Delta$ to
become large, up to order $1/m_b^3$ corrections; thereby achieving a
reliable interpolation between the two phase-space regions.

\section{Theoretical uncertainties}

Predictions for partial semileptonic decay rates can be broken up into
the individual contributions from different powers according to
\begin{eqnarray}
\Gamma_u &=& \Gamma_u^{(0)} + (\Gamma_u^{\rm kin(1)} + \Gamma_u^{\rm hadr(1)})
\nonumber \\
&& \mbox{} \hspace{6mm} + (\Gamma_u^{\rm kin(2)} + \Gamma_u^{\rm hadr(2)}) 
+ \ldots \nonumber \;,
\end{eqnarray}
where the superscript indicates both the nature of the correction and
their order in power counting in the shape-function region. For the
kinematical corrections $\Gamma_u^{\rm kin(n)}$ the sum of all terms is
known and can be used without truncating the series. We discuss the
uncertainty estimates term by term in the above formula.

The leading-power contribution has been factorized, as stated in
(\ref{eq:LPfact}), into hard, jet, and shape function. As mentioned
earlier, the leading shape-function uncertainty is coming from
experimental limitations, and is not accounted for in the list of
theoretical uncertainties. The hard and jet function have perturbative
expansions in the strong coupling constant $\alpha_s$, evaluated at
the scale $\mu_h$ and $\mu_i$, respectively. While $\Gamma_u^{(0)}$ is
formally independent of the choice of these scales, a residual scale
dependence is introduced by truncating the perturbative series. A
variation of the hard scale $\mu_h$ is used to estimate the unknown
2-loop contribution to the hard function. The uncertainty in the jet
function has been estimated by assigning $\pm [\alpha_s(\mu_i)/\pi]^2$
as a relative error. Here no scale variation has been used since a
change in $\mu_i$ also changes the shape function $\hat
S(\hat\omega,\mu_i)$, which is assumed to be extracted at a fixed
scale, which is chosen to be $\mu_i = 1.5$ GeV. However, the
uncertainty from the intermediate scale is somewhat outdated today,
ever since the appearance of the recent 2-loop calculation of the jet
function \cite{Becher:2006qw}.

The kinematical corrections $\Gamma_u^{\rm kin}$ start at order
$\alpha_s$ and are convoluted with the leading shape function. They
can thus be viewed as the product of subleading hard and jet
functions. However, for them the factorized expressions have not been
worked out, and the effects from the scales $\mu_h$ and $\mu_i$ have
not been disentangled. Instead, the expressions are evaluated at one
common scale $\bar\mu$ which is independent of, but around the
intermediate scale. Higher-order effects are again estimated by a
variation of $\bar\mu$.

So far we have accounted for three different perturbative
uncertainties resulting from the scales $\mu_h, \mu_i, \bar\mu$, which
can be added in quadrature. Next, we turn the discussion toward the
hadronic corrections. It is necessary to obtain an idea about the
severeness of our ignorance of the functional form of subleading shape
functions. To make even a central-value prediction we need to model
$\hat t(\hat\omega), \hat u(\hat\omega), \hat v(\hat\omega)$ while
respecting the moment constraints from OPE calculations. The idea for
an estimator for the associated uncertainty is to find many different
realistic models and study the resulting deviations from the default
value. This can be achieved by altering the functional forms of each
of the subleading shape functions via additional functions that have
vanishing first few moments. Using four such functions and either
adding or subtracting them, or leaving the subleading shape function
unchanged, already gives a set of $(2\cdot 4+1)^3 = 729$ different
models for the set of subleading shape functions. To be on the
conservative side, we use the maximal deviation from the central value
as an estimator for the subleading shape-function uncertainty. 

At order $(1/m_b)^2$ we expect to find many new shape functions. As
described earlier, we only include those contributions that feed into
the $(1/m_b)^2$ terms of the OPE result and neglect even smaller
contributions. At this level we need not worry about their precise
form and may model them using $(\mu_\pi^2/m_b^2)\,\hat S(\hat\omega)$
and $(\lambda_2/m_b^2)\,\hat S(\hat\omega)$. The differences between
these expressions and their true functional forms can be absorbed into
the subleading shape-function uncertainty.

Lastly there is one non-negligible error estimate at third and higher
order in power counting, which is the weak annihilation effect. This
error must be included whenever the experimental cut includes the
phase-space region near the origin, where $P_+ \sim P_- \sim
\Lambda_{\rm QCD}$. A recent study has put a limit of $\pm 1.8 \%$ on
the total rate by analyzing CLEO data \cite{TOMeyer}. A second
possibility is to cut out this region in phase space, at the moderate
cost (few events are located in that region) of a smaller
efficiency. We have found that this might improve the overall error
estimate slightly. 

In summary, we may split the theoretical uncertainty into three
categories: perturbative, hadronic, and weak annihilation. The sizes
of the individual errors in each of the categories depend on the
specific cut employed in the measurement. For example, cutting on
large charged-lepton energy is typically low in efficiency and is
significantly sensitive to subleading shape functions. In this example
the weak annihilation and hadronic errors dominate over the
perturbative one unless the cut is relaxed below $2.1$ GeV. (The charm
background starts at $2.3$ GeV.)  On the other hand, for more
efficient cuts, like a cut on the hadronic invariant mass or on $P_+$
for example, the perturbative uncertainty is typically dominant. At
present, the combined theoretical error on $|V_{ub}|$ -- excluding the
uncertainty from the leading shape function -- is in the neighborhood
of 5\%, but may be larger for certain cuts.

\section{Comment on ``Dressed Gluon Exponentiation''}

A different approach in calculating inclusive $B$-decay spectra
recently surfaced, named Dressed Gluon Exponentiation (DGE) by
Andersen and Gardi \cite{Andersen:2005bj,Andersen:2005mj}. This
computation is based on perturbation theory using an on-shell 
$b$-quark state ``dressed'' with gluons instead of the hadronic
$B$-meson state. An interesting consequence of this state is that the
kinematic range extends beyond the phase space of a single on-shell
$b$-quark state. This observation motivates to use the calculation in
comparing it with real data, so as to judge how realistic a dressed
heavy quark can mimic a heavy meson.

The procedure of dressing the $b$-quark with gluons requires a
complete knowledge of the anomalous dimensions of the jet and soft
functions (i.e.\ to all orders in $\alpha_s$), essentially because of
renormalization-group running deep into the non-perturbative
regime. Some limited information is available via the perturbative
calculation to first few orders, but the remaining aspects must be
modeled. Such models are guided by the large-$\beta_0$ approximation
and certain renormalon cancellations.

For these reasons the DGE calculation should be viewed as a model, and
not as a rigorous QCD prediction, the latter being a systematic
expansion and model-independent. While it might be interesting to test
this model against experimental data, it would be dangerous to use it
for the extraction of $|V_{ub}|$, since the uncertainty introduced by
the underlying assumptions are not under control. The statement that
no non-perturbative function is needed and that the ``prediction for
the spectrum depends {\em only} on $\alpha_s$ and on the quark
short-distance mass'' \cite{Andersen:2005bj} (emphasis as in that
reference) is not quite accurate; the same could be said about a
simple one-parameter model for the shape function, where the model
dependence is uncontrolled.

Lastly it should be noted that the DGE calculation has not been
performed to a comprehensive level comparable to the one described in
the previous sections, i.e.\ many power corrections that we addressed
above have yet to find their way into the DGE framework.

\section{Model-independent relations}

The universality of the leading shape function allows for infra-red
safe relations between different inclusive $B$-decay spectra. In fact,
the program outlined so far -- using the experimental data on the
$\bar B\to X_s\gamma$ photon spectrum to extract the leading shape
function, and subsequently plugging this function into the formula for
the semileptonic differential decay rate -- can be viewed as a
``manually shape-function free relation''. However, a more direct
relation is desirable because the extraction of the shape function is
somewhat cumbersome. Such relations would eliminate the issues arising
from the parameterizing the shape function, and different fitting
procedures that avoid sensitivity to resonances. The idea is to
``invert'' the jet function acting on the shape function in
(\ref{eq:LPfact2}) so that instead of convolving with the shape
function (which is not calculable) one convolves with the photon
spectrum (which is measured). The desired relation reads
\begin{eqnarray} \label{eq:SFfree}
\Gamma_u \Big|_{\rm cut} &=& |V_{ub}|^2 \int_0^{\Delta} dP_+\;
  W(\Delta_,P_+) \; \frac{1}{\Gamma_s} \frac{d\Gamma_s}{dP_+} \nonumber \\
&&\mbox{} + |V_{ub}|^2\;\Gamma_{\rm rhc} \Big|_{\rm cut}\;,
\end{eqnarray}
where the second term $\Gamma_{\rm rhc}$ collects residual power
corrections, e.g.\ terms that arise from the fact that different
combinations of subleading shape functions appear in the semileptonic
and radiative decay rates. The left-hand side of the equation denotes
the partial semileptonic decay rate as obtained after imposing a
cut. For the determination of $|V_{ub}|$ the partial $\bar B\to
X_u\,l^-\bar\nu$ decay rate, as well as the photon energy spectrum,
enter as experimental inputs. The weight function $W(\Delta_,P_+)$ and
the residual terms $\Gamma_{\rm rhc}$, on the other hand, are
theoretical quantities.

The main obstacle in finding an infra-red safe weight function is that
the jet function $J(p^2)$ in (\ref{eq:LPfact2}) is a distribution,
which does not ``invert'' easily. Although the jet function is
universal, it cannot be eliminated entirely either, because it is
called with different arguments in (\ref{eq:LPfact}) and in
(\ref{eq:LPfact2}). The difference is that in radiative decays $y=1$,
while this is generally not true for semileptonic decays.  This
problem is solved by defining a ``jet kernel'' $Y(k,\ln y)$ which is
used to extract the $y$-dependence from the jet function by means of
the following equation \cite{Lange:2005xz}
\begin{equation}
\int\limits_0^{y m_b \Omega} dp^2\, J(p^2) = \int\limits_0^\Omega dk\,
Y(k,\ln y) \int\limits_0^{m_b (\Omega-k)} dp^2\, J(p^2)\;.
\end{equation}
This equation defines the jet kernel to all orders in perturbation
theory. The explicit expressions for $Y(k,\ln y)$ have been calculated
to complete 2-loop order. We now proceed in constructing the weight
function.

The experimental cut on the $\bar B\to X_u\,l^-\bar\nu$ phase space
can be encoded as
\begin{eqnarray}
 0 &\le& P_+ \le \Delta \;, \nonumber \\
 0 &\le& y \le y_{\rm max}(P_+) \;, \\
 0 &\le& \e \le \e_{\rm max}(P_+,y)\;. \nonumber 
\end{eqnarray}
As written in (\ref{eq:SFfree}) the weight function depends on
$\Delta$. More generally, the weight function (and also the power
correction $\Gamma_{\rm rhc}$) changes as the specifics of the cut
change, that is, it depends also on the functions $y_{\rm max}(P_+)$
and $\e_{\rm max}(P_+,y)$ that encode the cut. When integrating the
leading-power differential decay rate formula (\ref{eq:LPfact}) over
this phase space one can show via simple interchanges of integrations
that the weight function $W^{(0)}(\Delta, P_+)$ itself factorizes
symbolically as
\begin{equation}
W^{(0)}(\Delta, P_+) \sim H_\Gamma \; H_u(y,\e) \otimes Y(k,\ln y)\;,
\end{equation}
where this time $\otimes$ is a three-folded convolution in the
variables $\e,y$ and $k$. Because the specifics of the cut, $\Delta,
y_{\rm max}(P_+)$, and $\e_{\rm max}(P_+,y)$ enter only as integration
limits in this structure, weight functions can be computed in an
automated fashion. 

Before we turn the discussion to the error analysis when using
relation~(\ref{eq:SFfree}) for the extraction of $|V_{ub}|$, we have
to note a feature that contrasts direct partial decay-rate prediction
described in Section~\ref{sec:2}. There, the $b$-quark mass enters
only indirectly into the factorization formula~(\ref{eq:LPfact})
through the shape function. When integrating the differential decay
rate to obtain a partial rate, the $m_b$ dependence is generated
dynamically. For example, the total decay rate is proportional to
$\int dP_+\,(M_B-P_+)^5 \hat S(P_+) \approx m_b^5$ at tree level. The
analogous statement is true for the $\bar B\to X_s\gamma$ photon
spectrum. This is the reason for the $m_b^3$ factor in the denominator
in (\ref{eq:LPfact2}) for the {\em normalized} photon
spectrum. Therefore the weight function $W(\Delta, P_+)$ will pick up
this explicit $m_b$ dependence. Previous works
\cite{Leibovich:1999xf,Leibovich:2000ey} used the {\em absolute}
photon spectrum, which introduces other, more severe
uncertainties. Firstly, without the normalization one can only extract
the ratio $|V_{ub}|^2/|V_{ts}^* V_{tb}|^2$. Secondly, the radiative
corrections are then unacceptably large due to large operator mixing
effects \cite{Lange:2005qn}. Thirdly, event fractions in $\bar B\to
X_s\gamma$ can be calculated with higher precision than the absolute
branching ratio \cite{Neubert:2004dd}.

As an example let us consider the theoretical uncertainties on the
partial decay rate when using relation~(\ref{eq:SFfree}) for a pure
cut on $P_+$. To this end we may pretend for the moment that the
experimental data on the shape of the photon spectrum had no
uncertainty. The analysis was carried out in \cite{Lange:2005qn}, and
reads
\begin{eqnarray}
&& \Gamma_u(P_+\le 0.65\;{\rm GeV}) \\
&=&
(46.5 \; \pm 1.4\,\hbox{\scriptsize [pert]}
        \; \pm 1.8 \,\hbox{\scriptsize [hadr]}
        \; \pm 1.8 \,\hbox{\scriptsize [$m_b$]} \nonumber \\
&& \mbox{} \; \pm 0.8 \,\hbox{\scriptsize [pars]}
        \; \pm 2.8 \,\hbox{\scriptsize [norm]} 
)\;|V_{ub}|^2 {\rm ps}^{-1} \nonumber 
\end{eqnarray}
Again, the perturbative uncertainty stems from a variation of the
renormalization scales, and the hadronic uncertainty from scanning
over many models of subleading shape functions. While the latter is
comparable to the error analysis of the direct next-to-leading order
partial rate prediction, the perturbative error is significantly
reduced due to the fact that the jet kernel has been evaluated to
complete 2-loop order. (In the direct prediction \cite{Lange:2005yw}
we find $\pm 2.5\, \hbox{\scriptsize [pert]}\;|V_{ub}|^2 {\rm
ps}^{-1}$ with the 1-loop jet function.) Varying the input parameters
needed for the weight function, i.e.\ quark masses and HQET
parameters, leads to the next two stated errors. Finally we have to
consider the norm of the relative photon spectrum, which impacts
$|V_{ub}|$ directly. Because the shape of the photon spectrum is not
measured over the full kinematic region but only above some photon
energy $E_0$, we will need the theoretical prediction for the event
fraction that falls into that window \cite{Neubert:2004dd}.

\section{Conclusions}

The CKM matrix element $|V_{ub}|$ is a fundamental parameter of the
Standard Model, which is already the first motivation in pursuing its
precise determination. It also allows for testing the CKM picture of
{$\cal CP$} violation and for indirect New Physics searches. A
determination with overall uncertainty around and below the 10\% level
is both experimentally and theoretically challenging, but feasible. In
this talk we have touched upon many of the theoretical issues and
possibilities, from direct predictions of partial rates using
QCD-factorization theorems to relations between different inclusive
$B$-decay spectra, which are model-independent at leading power.

The theoretical error can be reduced in several ways. Higher-order
computations allow for a reduction of perturbative uncertainties. The
recently published 2-loop result for the jet function
\cite{Becher:2006qw} will have an impact on both the direct
theoretical prediction of partial rates -- via reduced scale
dependence -- and the weight-function relations. While the weight
function itself was already known to 2-loop order at the intermediate
scale, the improved jet function will impact the theoretical error on
the event fraction in $\bar B\to X_s\gamma$, i.e.\ the {\em norm} of
the photon spectrum. Other possibilities for improving the theoretical
errors include the computation of higher-order hard functions;
however, a complete next-to-next-to leading order description in
renormalization-group improved perturbation theory would require
3-loop anomalous dimensions and the 4-loop cusp anomalous dimension,
which seems currently out of reach.  On the non-perturbative side,
further studies of subleading shape function contributions, and
improved limits on weak annihilation can have a similar overall
impact.

A precise determination of $|V_{ub}|$ requires a variety of different
methods and measurements. The study of inclusive $B$ decays remains a
very active field of research with further improvements in both theory
and experiment.

\bigskip 
\begin{acknowledgments}
I would like to thank the organizers of the FPCP2006 conference for a
very enjoyable time in Vancouver. I also thank Stefan Bosch, Matthias
Neubert, and Gil Paz for their collaboration on much of the work
reported here, and Iain Stewart and Zoltan Ligeti for
discussions. This work was supported in part by funds provided by the
U.S.~Department of Energy (D.O.E.) under cooperative research
agreement DE-FC02-94ER40818.
\end{acknowledgments}

\bigskip 

\end{document}